\documentclass[conference]{IEEEtran}
\IEEEoverridecommandlockouts
\usepackage{cite}
\usepackage{amsmath,amssymb,amsfonts}
\usepackage{algorithmic}
\usepackage{graphicx}
\usepackage{textcomp}
\usepackage{xcolor}
\usepackage[tableposition=top]{caption}
\usepackage{placeins}
\def\BibTeX{{\rm B\kern-.05em{\sc i\kern-.025em b}\kern-.08em
    T\kern-.1667em\lower.7ex\hbox{E}\kern-.125emX}}
\begin{document}

\title{Heartbeat Sound Classification with Visual Domain Deep Neural Networks\\

}

\author{\IEEEauthorblockN{1\textsuperscript{st} Uddipan Mukherjee}
\IEEEauthorblockA{\textit{School of Computer Science and Mathematics} \\
\textit{Liverpool John Moores University}\\
Liverpool, UK \\
mukherjeeuddipan@ieee.org}
\and
\IEEEauthorblockN{2\textsuperscript{nd} Sidharth Pancholi}
\IEEEauthorblockA{\textit{Weldon School of Biomedical Engineering} \\
\textit{PURDUE UNIVERSITY}\\
Delhi, India \\
s.pancholi@ieee.org}
}

\maketitle

\begin{abstract}
Heart disease is the most common reason for human mortality that causes almost one-third of deaths throughout the world. Detecting the disease early increases the chances of survival of the patient and there are several ways a sign of heart disease can be detected early. One of easiest and convenient way to identify cardiovascular disease is heart sound auscultation technique by digital stethoscopes and mobile device applications. Though data can be collected electronically using these methods, identification of any problems in the mechanical heart sound is still a manual process and requires expert clinical personnel to diagnose any abnormalities. Thus, research on automated feature extraction and classification of cardiovascular sounds is necessary. Pascal heart sound classification challenge dataset has been popular with previous researches as it provides a good variety of cardiovascular audio with consistent categories. This study proposes to convert cleansed and normalized heart sound into visual mel scale spectrograms and then using visual domain transfer learning approaches to automatically extract features and categorize between heart sounds. The study will use visual domain classification approaches i.e convolution neural network-based architectures i.e. ResNet, MobileNetV2, etc as the automated feature extractors from spectrograms. These well-accepted models in the image domain were found to learn generalized feature representations of cardiac sounds collected from different environments with varying amplitude and noise levels. Model evaluation criteria used were categorical accuracy, precision, recall, and AUROC as the chosen dataset is unbalanced. The proposed approach has been implemented on both datasets of the PASCAL heart sound and resulted in \~ 90\% categorical accuracy and AUROC of ~0.97 for both sets.
\end{abstract}


\section{Introduction}
Heart disease can be of different types i.e. blockage in an artery, heart valve, muscle problems, and rhythmic issues. Detecting at early stages holds a lot of significance for effective treatment of heart disease. Both invasive and non-invasive diagnosis methods are available which produce dependable results. Some of the major invasive methods to test cardiac heath are blood tests, coronary angiograms, etc. Non-invasive methods include an exercise stress test, electrocardiogram, auscultation techniques, magnetic resonance imaging, heart rate monitoring, chest x-ray, etc. The heart sound auscultation technique is the easiest, convenient, inexpensive, and quick way to understand the heart health among the discussed, but it requires clinical expertise. This technique is the oldest of any other cardiovascular health check and can be traced back to the 19th Century. The mechanical motion of the cardiovascular system produces informative sounds that contain details of the interplay between different sections of the heart. The presence of any abnormality such as noise, warp sound can indicate heart disease at early stages.

 Several heart-related irregularities can be found using simple and convenient techniques like the usage of stethoscopes, mobile applications, etc. The cardiac sound is a feeble signal that gets generated by opening and closing of tricuspid, mitral, aortic, pulmonic valves, quivering of myocardial contraction and relaxation, and bloodstream impacts. The heart generates mechanical audible sounds i.e., lub and dub (Clinically termed as S1 and S2). During normal circumstances S1 prolongs for 50 – 100 milliseconds with a frequency of 50 – 150 Hz, whereas S2 clocks 25 – 50 milliseconds with 50 – 200 Hz frequency. The occurrence pattern of these sounds makes out of different categories i.e. Normal, Murmur, Extra Sound, ExtraSystole. Murmur sound mostly indicates illness and is graded between 1 to 6 depending upon the loudness. Extra sound or Extra Systole types may or may not indicate disease. Knowing the type of heart sounds helps doctors to decide on the course of action for the patient. Identification of these sound categories remains manual till date for most cases. Also, one of the major disadvantages of the cardiac sound auscultation method is background noise and external inferences. This can cause diagnostic issues in case the measurement has been done by an inexperienced clinical person. Most of the previous works focused on manual feature extraction from audio frequencies i.e. time, frequency, amplitude, etc. Once features are extracted, they are sent as input to classification learning deterministic or probabilistic algorithms. The generated models then were able to successfully predict several classes of heart sound with moderately well accuracy and precision.

\cite{b8} worked on denoising the heart audio with wavelet and short-time Fourier transformation in a combination to denoise the audio signals after a resampling step. In the feature extraction step, the study extracted statistical features of heartbeat, systole, and diastole i.e. sum, mean, standard deviation, etc. along with the position of S1 and S2 sound (Lub and Dub). Conditional logic has been used to classify the extracted features and results were measured by precision, F-Score, Youden’s Index, and Discriminant Power, but did not result in a good accuracy. \cite{b28} focused on the classification of Normal vs Murmur heart sound based on audio time-frequency and non-linear feature extraction. Noise was eliminated using power line de-noising and white noise removal methods. Five distinguishable features based on time-frequency and non-linear entropy methods were extracted from the cleansed signal. SVM with Gaussian Radial Basis kernel classifier was then used to categorize both types that resulted in ~97\% accuracy. This was not stable as heart sounds with different frequencies resulted in prediction errors as wavelet decomposition methods were used. \cite{b7} proposed a novel method in their study where auto-correlated features were used with a diffusion map for low dimensional feature extraction. SVM reached a sensitivity of ~38\% in the case of problematic heart sounds in both Murmur and Extra systole categories, which are inferior to other studies. \cite{b27} focused on heart sound segmentation and then the creation of a scaled spectrogram from the segmented heart cycles. Assignment of S1 and S2 sound was achieved on the consideration that the diastolic period is more than the systolic period. Log power spectrogram of the estimated heart cycle is then created using a Short-time Fourier transformation. Tensor decomposition is then used to extract discriminative features from the scaled spectrograms. SVM classifier further categorized the sounds with a good precision score. \cite{b21} analyzed heartbeat signals visually by plotting the amplitude vs time plots and spectrograms which showed that the different categories of heart audio are visually distinguishable with help visual representations. The study generated fixed time length signals of 50000 frames from the denoised dataset for classification. The best-performed algorithm was found to be long short term Memory (LSTM) with regularizations. Accuracy was recorded around 80\% with five-fold cross-validation.

There has been some work done on the implementation of visual image models with the environmental sound dataset. Recent studies have showcased promising results while trying to classify audio domain datasets using visual domain models like CNN, CRNN, etc. \cite{b11} worked with the classification of the GTZAN dataset which includes music from 10 genres. \cite{b5} explored spectrograms generated with varied parameters along with several deep convolutional neural networks for automated tagging of music. The study concluded that the mel-spectrogram to be a better representation than raw spectrograms for automated audio classification problems, and deeper networks can result in better accuracy scores when more training data is available. \cite{b3}) has experimented with spectrograms, MFCC (Mel-Frequency Cepstral Coefficients), CRPs (Cross Recurrence Plots) representations to find out the effectiveness of each type and concluded that spectrogram is the most effective visual representation which can be classified with high accuracy by visual domain deep models (i.e. GoogleNet, ResNet, etc). Usage of temporal information in audio classification was explored by \cite{b16} where mel-spectrogram was used as input to Convolutional Recurrent Neural Network (CRNN). \cite{b18} considered using pre-trained Imagenet models to classify well-known audio data sets like GTZAN, ESC-50, and UrbanSound8K. Variable window size and hop lengths were used to calculate the three-channel Mel-Spectrogram to capture a variable level of frequency and time information. Both randomly initiated and pre-trained models were used for the experiment. It was observed that the  pre-trained models had outperformed randomly initialized models. The study also showed that the weights did not change much for first few layers of pre-trained models, but weights in middle layers went through significant changes to accommodate spectrogram-related features.

Convolution neural networks are known to be working best with a large quantity of input training datasets. This helps the network to generalize better on the unseen dataset by learning complex nonlinear functions that in turn results in better accuracy and also help to avoid overfitting problems. Heart sound databases lack quality tagged training data when compared with image domain datasets. One of the easy and effective solutions to this problem is data augmentation. \cite{b22} explored several audio augmentation techniques on environmental sound datasets and their impact on the overall accuracy score. The techniques experimented with were time stretching, pitch shifting, dynamic range compression, and addition of background noise. All these augmentations were applied directly to the audio before transforming to mel-spectrogram. Models have shown ~16\% gain in accuracy wherever data augmentations are included. Further analysis of the results found that, though augmentation overall improved accuracy per class for some classes it was reduced due to some overlapping characteristics. Hence, the study proposed class-wise augmentation based on these findings for future experiments. As the generated spectrogram is an image representation, several image augmentation techniques can be used as well to reduce overfitting. Some common image augmentation techniques include random rotations, random crops, horizontal flipping, and affine transformations. \cite{b9} introduced a new approach that they called CutOut. This technique randomly masks out square regions in an image to augment it. The application of this approach to the input dataset improved the performance and robustness of related CNNs. Similar to CutOut, \cite{b29} have explored another approach by erasing a rectangular space in an image. This method was named random erasing, it was proved to be a useful method to avoid overfitting and improve the generalization capability of CNN-based models. Another effective image augmentation method SamplePairing was explored by \cite{b15}, the method consolidates one image with another randomly chosen image from the training dataset. The method which randomly combines the pixels of two images to generate a sample image is called MixUp, which was explored by \cite{b26}. The point to be noted here is that SamplePairing is a special case of the MixUp method where the ratio of the pixel mix becomes 1:1. Both approaches showcased notable advances in the classification accuracy of image domain models. \cite{b20} at Google proposed a spectrogram augmentation technique termed SpecAugment that is simple and computationally cheap. The approach includes time wrapping, time and frequency masking-based augmentations. Time wrapping refers to warping the X-axis (time) of the spectrogram time series. Time masking refers to masking a group of successive time steps whereas frequency masking refers to masking frequency channels in a spectrogram. The approach had been tested with speech recognition problems with ASR (Automatic Speech Recognition) networks and obtains the state of the art results for LibriSpeech 960h by \cite{b19} and Switchboard 300h by \cite{b10} tasks. One of the major points noted during the study was that the time wrapping did help to improve the performance of the model but was not a crucial contributor. It also noted that augmentation converted an overfitting problem into an underfitting one, which was then solved by common methods like designing longer networks or training for longer duration etc.

There has been significant research on visual domain data such as images and videos during the last decade which contributed several state-of-the-art models for feature extractions. The ImageNet challenge required these models to be trained on a huge data set (1 Million +). These trained models were able to learn image data representation layer by layer. Initial layers learned simple features like edges, boundaries whereas later layers learned more complex features like the species, background, environment, etc. This knowledge can be readily used as network weights and biases for feature extraction of any similar images. Literature review found out that none of the studies has investigated automated feature extraction using visual methods of cardiac sounds yet. A near real time and automated feature extractor along with a classifier of cardiac audio can be helpful for the medical domain and health care industry, that can incorporate the method to electronic auscultation devices for detection of heart sound categories. 

This paper makes the following contributions:

\begin{itemize}
\item	Successfully demonstrates how visual domain neural networks can be used to categorize heart audio data with good accuracy. The method described in this study can be further used to train on other cardiovascular datasets and predict high-level categories. 
\item   Exhibit training benefit in terms of normalization, data availability and clipping out invalid information by splitting a curated heart sound into multiple audio snippets of standardized duration. 
\item	Classification performance comparison between different pre-trained feature extractors i.e ResNet152V2, MobileNetV2, etc. based on the training/testing performed with the Pascal heart sound classification challenge dataset.
\item	State of the art spectrogram augmentation method SpecAugment experimented with speech data and showed good speech recognition results. As per our knowledge, this study has used the SpecAugment method to cardiovascular audio data for the first time. Implementation of the method, augmented the heart sound spectrograms effectively by time wrapping and frequency masking techniques and boosted classification accuracy.

\end{itemize}

The paper is organized as follows, research methodology along with dataset details are described in section II. This section also focuses on the actual implementation of the end-to-end classification process including details on parameter and hyper-parameter choices, processes, and libraries. Results and analysis of the study are captured in section III. Section IV and V wraps up the research, summarizes the new information, makes recommendations, and suggests further research in the field.

\section{RESEARCH METHODOLOGY}

Audio data interpretation by manual visual techniques has been widely accepted in different domains and use cases. This encouraged the idea of experiments with visual domain models with the audio dataset, and some of the previous works have shown significant potential with the approach. 

\subsection{Dataset Details}

The PASCAL heart sound challenge created by \cite{b30} has captured heart sound data from digital stethoscopes and an iMobile application named iStethoscope Pro. Labeled and unlabeled raw heart audio files with variable length between 1 to 30 seconds are available in both datasets. Though the unlabeled dataset can be classified by the final model, it cannot be used for performance measurement. Hence the unlabeled dataset is excluded from this study. Data collected by the iStethoscope Pro (Known as Dataset-A) application has four categories, i.e. Normal, Murmur, Extra Heart Sound, \& Artifact. Dataset-A has a total of 124 labeled and 52 un-labeled audio files in WAV format. As this data was collected via mobile devices, environmental noise can be predominantly found as the primary type of noise. The iPhone microphone requires it to be pressed to the apex region of the heart on the direct skin to capture the heart audio with a 44.1 kHz sample frequency. Dataset-B was collected from hospitals by digital stethoscope and is categorized into three classes, i.e. Normal, Murmur \& Extrasystole. This has 461 labeled and 195 un-labeled audio files in WAV format. The labeled dataset is unbalanced where ~70\% of audio files belong to the Normal category only. Digital stethoscopes recorded heartbeat audio sound with a 4 kHz sampling frequency. Most of the noise in this dataset can be attributed to breathing, lung sounds, and electronic inference. Actual heart audio on both datasets is mostly available in the lower region of the frequency scale whereas the higher frequency region contained noise information generally. Dataset description mentions using a low pass filter at 195 Hz to extract the information related to clean heart sound. Dataset available in each category has intra-class similarities and inter-class variations which is ideal for classification problems.

The PASCAL heart sound challenge presented two problems. First to identify the locations of lub(S1) and dub (S2) sound via segmentation techniques. Some of the audio files contained the exact location of S1 and S2 sounds which needed to be used for learning the location with the help of machine learning techniques. The current research aims to solve the second challenge that requires the heart sounds to be categorized directly with appropriate labels without identifying the location of lub and dub sound. Datasets A and B do not consist of the same categories and capturing method, environment, the sampling rate was different. Hence, combining the datasets is not recommended for automated feature extraction and classification. Dataset A has fewer data per class, but it seems to be balanced. Dataset B has adequate training data, but it reflects class imbalance problems where normal category ~ 3 times that murmur and ~ 8 times than extrasystole category. The normal category of both the dataset indicates conventional and robust heart sounds. It will have a clear lub-dub pattern with some surrounding noise. From normal heart sounds it can be noted that the diastolic period is longer than the systolic period. The normal heart rate per minute is between 60 and 100 when the person is at rest. The dataset captured both cardiac audio patterns from children, adults at both excited and rest states. Murmur sound is one of the indicators of several disorders of the heart. It can be identified as a complication based on the presence of other factors i.e. chronic cough, chest pain, etc.

A murmur can be either systolic or diastolic. It happens when the flow of blood between the heart chambers is not regular, which creates noise during the systolic or diastolic period. Murmur sound appears between S1 and S2 sounds and rarely overlaps during the actual cardiac sounds. The regular appearance of an added lub or dub sound at the end of S1 or S2 respectively can be categorized as an Extra heart sound. It sounds like “lub-lub dub” or “lub dub-dub”. It can be an indicator of disease sometimes but can also be a normal condition. Extrasystole pattern of heart sound includes random addition or absence of either lub or dub sounds. Point to be noted as extra-systole sounds are added or removed randomly as opposed to regular additional sounds in the Extra heart sound category. Extrasystole sound may or may not be a symptom of an ailment. Extrasystole is frequently found in children and can also be found in adults sometimes. The visual analysis of the data helps to point out the difference between each category. Figure 1 shows the pattern difference between different heart sounds.

Artifact sound does not include any audible heart sound, but only background noise. These kinds of sounds can be captured by mobile devices as the background environments are not stable. Most of the scenarios have a frequency greater than 195 Hz that indicates only noise information. Features of this category should show different characteristics than any other categories with cardiac sound. The artifact dataset does not provide much temporal information on cardiac sounds.

\subsection{Data Preprocessing}

Chosen variable length dataset of the study has audio noise, invalid information, and class imbalance. A common way to achieve denoising is to use a low pass filter that omits audio frequency above recommended 195 Hz to eliminate most of the background noise. Ideally, Fast Fourier Transform can be applied to the signal for identifying the frequency and the amplitude components of noise vs the original signal. As the cutoff frequency is already present, the study will not incorporate complex methods like multi-level SVD or compressed sensing to denoise the dataset. Low pass or high cut filters are designed to pass a low-frequency range of choice while rejecting the higher ranges. A simple low pass filter with a capacitor and resistor gets used for this purpose traditionally, R is the value (Ohms) of a resistor, C is the capacitor, V\_inis the input voltage and V\_outis the output voltage.

\begin{equation}
\frac{d V_{o u t}}{d t}+\frac{1}{R C} V_{o u t}=\frac{1}{R C} V_{i n}
\end{equation}

This study will implement the low pass filter programmatically using python Scipy signals libraries. Firstly, Nyquist frequency was calculated with the help of extracted sampling rate from the audio signal. Then the calculated frequency was used to determine the normalized cutoff frequency. Digital butterworth filter was implemented to calculate the numerator and denominator polynomials of the infinite impulse response filter. Zero phase filters were then used along with the extracted numerator, denominator polynomials, and the audio signal to calculate the new denoised signal. Dataset A captured by an iMobile mobile application, recorded the heart audio sound with a sampling frequency of 44,100 Hz. Processing a signal with such a high sampling rate would require more processing power,resources and time, hence the study down-sampled dataset A by a factor of 10. Digital resampling of the signals was achieved with help of the librosa python library and took place just before the denoising action.

Audio length needs to be normalized as length is different of each available audio file in the dataset. This study proposes to split each record by 3 seconds duration to standardize the audio length. The reminder part if any from the tail was discarded. This strategy also helped to eliminate stethoscope-related disconnection noise with dataset B. Records with audio length less than 3 seconds were ignored with the implementation of the split method. Python library “wave” was used for the split implementation and the output files are stored for further reference. Dataset A does not have a large database of cardiac sounds which can result in generalization issues. Hence it is important to augment the data so that each category does not lose out on its basic characteristics. This study proposes to use the time and pitch shifting based audio augmentation techniques for primary augmentation. Time-shift modifies the tempo of an audio signal without updating the pitch. Pitch shifting techniques help for heart audio classification as it alters the frequency component of the sound itself. Pitch shift implemented with python librosa library by shifting down the pitch by one-half step. Right time shift was applied to all the categories of the dataset by 1 unit. 

\begin{figure*}[htbp]
{\includegraphics[width=500pt]{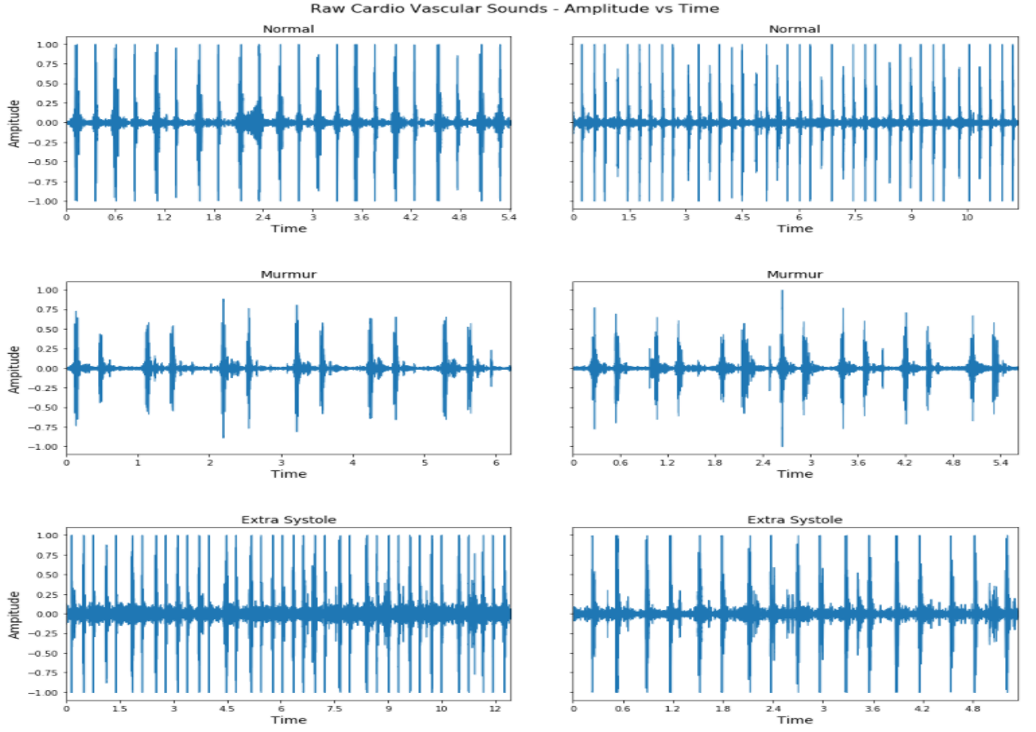}}
\caption{Amplitude vs Time Plots of cardiovascular sounds}
\label{fig}
\end{figure*}

Different values experimented for the pitch and time-shift where one unit resulted in the best values. Both the applied transformations resulted in varied samples of data for all the categories. The augmentations were implemented only with the training data of dataset A. Dataset B already has a considerable amount of audio files after the split; hence the audio augmentation techniques were not implemented with it.

\subsection{Transformation into Image}
There are several ways to convert audio in a spectrogram representation, i.e. Fourier Transform, Filter-banks, etc. Several studies i.e. \cite{b5}, \cite{b1} have shown that spectrograms with mel scale perform better with visual domain models. Keeping this as a note, this study will use the short-time Fourier transform method to generate log/mel scale spectrograms. As a first step, a short-time Fourier transform requires to be calculated with the help of equation 2.

\begin{equation}
X(\tau, \omega)=\sum_{n=-\infty}^{\infty} x[n] w[n-\tau] e-j \omega n
\end{equation}

$ \omega $ is the phase of the basis sinusoidal, $ \tau $ is time points in a time-domain signal x. Calculating equation 2 requires the raw signal to be divided into overlapping frames which are then multiplied by a window function w. The same length of the Fast Fourier Transform window was used for both the dataset with a value of 200. The window function is required to be applied to lower spectrum disturbances created by the framing. Generated magnitudes of STFT to be plotted which can be termed as magnitude spectrograms. Transforming this to mel-spectrogram requires frequencies to be warped to the mel-scale, and fast Fourier transform bins to be merged to mel-frequency bins. Then the Mel-scaled power-spectrogram is obtained by multiplying the squared magnitude-spectrograms with the mel-filter-bank. At last, the log power spectrogram or mel-spectrogram can be calculated with help of Equation 3.

\begin{equation}
S=10 \log 10|X(\tau, \omega)| 2
\end{equation}

Mel-spectrograms generated by the above method are then plotted as RGB images to visualize the change of frequency over time and varied amplitude with 8000 as the maximum value for mel-frequency scales of the Y-axis. The decibel scale mel-spectrogram was then converted to a power spectrogram for visualization with help of the librosa python library. The study intends to perform spectrogram augmentation techniques to increase data availability for training and validation which will help the classification to be generalizable. A voice audio spectrogram augmentation technique named SpecAugment was proposed by \cite{b20}. Our experiment aims to extend SpecAugment implementation to cardiac sound datasets. The technique offers time wrapped and frequency block masked spectrograms. Spectrogram augmentation to be directly applied only on the denoised and normalized raw audio dataset and not on audio augmented samples. For Dataset A, one time-wrapped and two frequency masked images were generated per each available spectrogram including the audio augmented dataset using SpecAugment. Each spectrogram image in dataset B was augmented with one time-wrapped and one frequency masked transformation. A simplistic implementation \cite{b24} of specAugment using NumPy and SciPy libraries was used in this study. Pixel values of all augmented colored spectrogram images are required to be scaled between 0 and 1 for normalization before they could be used with deep learning models.  Pixel values of the images used by the study vary between 255 to 1, hence simply dividing the values by 255 scaled the values to the required range. Normalized images were then re-sized to fit the memory and batch capacity required by the model training. It was found the resizing all spectrogram images to 128*128 dimensions generate the best results with the capacity of the hardware used during the training process. Numerical versions of the images were then shuffled across to meet IID (Independent and identically distributed) assumption.

\subsection{Data split and class balancing}

The study suggests train and test data split with a 70:30 ratio. The best practices recommend using a validation dataset to ensure a good performing model. As the size of available dataset is already less, splitting it further for validation will result in an insufficient training dataset. Augmentation techniques to be used for generating more datasets from the training set.Augmented data can then be used in an 80:20 split, where 20\% of the augmented data to be used for validation purposes. Augmentation techniques were not used with the testing dataset.

Dataset A has a similar count of murmur and normal samples which is around half of the occurrence of the artifact category. A down-sampling method was applied to the artifact class where 50\% of the dataset belonging to the class has been randomly eliminated. The technique balanced artifact, normal, and murmur classes. Class imbalance persisted with extrahls class, though the method increased the weight value of the class. Dataset B showcases a similar trend where 62\% of the overall set belongs to the “normal” class and the lowest frequency “extra sound” class consists of only 8\% of the dataset. The murmur class in dataset B has a frequency of 30\%. To balance the normal class with the murmur class, a significant number of samples were required to be removed which could impact the data distribution of the class. Due to this uneven class imbalance, down-sampling or up-sampling of any class may not work effectively, hence the technique was not applied to dataset B. Another known and effective technique to handle such imbalance is known as class weighting. This approach determines the weight of each class and assigns the calculated value while training. Using class weights parameters while training changes the range of the loss. The loss becomes a weighted average and helps to tune SoftMax results accordingly. Weight for each class was calculated as the number of occurrences of the class divided by the number of records available in the training dataset. To conclude, dataset A included both down-sampling and class weighting to handle class imbalance whereas dataset B included only the class weighting technique for the purpose. 

\subsection{Interpretation/Evaluation}

As a multiclass classification problem, some of the suitable measures for model evaluation are categorical accuracy, precision, recall, specificity, AUROC, etc. Precision and recall focus on True Positives and True Negatives respectively, hence, helps to generate insights with an imbalanced dataset. Area under ROC measure shows capability of the model to discriminate between classes. AUROC score is calculated with help of precision and recall, hence works as a robust metric for the imbalanced dataset as well. Equations for precision, recall, and specificity has been shown in equations \# 4,5,6 respectively. 

\begin{equation}
\text { Precision }=\frac{\text { True Positive }}{\text { True Positive+False Positive }}
\end{equation}                 
\begin{equation}
\text { Recall }=\frac{\text { True Positive }}{\text { True Positive }+\text { False } \text { Negative }}
\end{equation}                                                      
\begin{equation}
\text { Specificity }=\frac{\text { True Negative }}{\text { False Positive }+\text { True Negative }}
\end{equation}

\subsection{Visual domain neural network}

Convolutional Neural Networks are advanced architectures that excel at processing visual data, such as images and videos. CNN's revolutionized classification tasks for the visual domain that were previously thought to be incredibly difficult. The architecture of CNN was inspired by the animal visual system and recently it has outperformed humans in classification accuracy. A common CNN architecture will consist of a few convolution blocks followed by pooling layers iteratively until the matrix size of the feature map gets small enough for the dense layer. The feature map is converted to a feature vector which is then sent as input to the dense layer and then the classification layer for categorization.

Though the common CNN architecture results in a good performance for some dataset, complex problems such as the ImageNet challenge requires more sophisticated network design. It was observed that stacking up more convolution, pooling layers resulted in better accuracy till a certain point, after that addition of more layers caused information loss. Some efficient CNN-based networks were compared in \cite{b4}.  The paper compared the accuracy, number of Operations, number of network parameters, training and test time, power requirement, Accuracy density per network parameter, etc. It showed models with better accuracy required a lot of computational time with the high volumne of model parameters. Hence a trade-off between the number of parameters and accuracy required to be set up. This study aims to classify cardiac sounds with good accuracy, but also will consider the practical use of the model. Future applications might include the cardiac sound classifier in real-time environments i.e. mobile devices; hence the prediction speed will be important. Hence the study needs to find a balance between a good accuracy and a smaller number of parameters. ResNet proposed by \cite{b12} and MobileNetV2 proposed by \cite{b23} seems to be two good architectures for this purpose with good ImageNet accuracy score. ResNet152V2 \cite{b13} and MobileNetV2 have ~60 million and ~3.5 million trainable parameters respectively.  The study will also include experiments with networks like MobileNet \cite{b14}, InceptionResNetV2 \cite{b25}, and Xception \cite{b6} for both datasets. Table I captures some data points about some CNN variant models.

\begin{table}[h]

\caption{Comparison of CNN variant architectures}

\begin{tabular}{|l|l|l|l|}
\hline
\textbf{\begin{tabular}[c]{@{}l@{}}CNN   Variants\\ /Parameters\end{tabular}}   & \textbf{\begin{tabular}[c]{@{}l@{}}MobileNet\\ V2\end{tabular}} & \textbf{Xception} & \textbf{\begin{tabular}[c]{@{}l@{}}ResNet\\ -152V2\end{tabular}} \\ \hline
\begin{tabular}[c]{@{}l@{}}\# of trainable parameters \\ (Million)\end{tabular} & $\sim$3.5                                                       & $\sim$22.9        & $\sim$60.3                                                       \\ \hline
ILSVRC Accuracy                                                                 & 0.901                                                           & 0.945             & 0.942                                                            \\ \hline
Model Size (MB)                                                                 & 14                                                              & 88                & 232                                                              \\ \hline
\# of Layers                                                                    & 88                                                              & 126               & 152                                                              \\ \hline
\end{tabular}
\end{table}

\subsection{ResNet-152V2}

Residual learning network (ResNet) is a variation of CNN which enables deeper network learning effectively. ResNet avoided the vanishing/exploding gradients problem of deeper networks with help of residual learning. In a residual network shortcut or skip connections can be inserted when input and output dimensions are the same as the stacked convolutional blocks. Skip connection adds the input of the first convolution block to the output of the next convolution block in a 2-layer stack. When the input and output size differ then shortcuts can either pad the output with zero or project to match dimensions with help of 1*1 convolution. This helps to reformulate the layers explicitly by identity mapping to address the degradation issues. Shortcut connections allow the gradient to flow through itself which helps to mitigate the vanishing gradient problem. The identity function of ResNet ensures that the subsequent layers learn at least like the previous layers if not better. The ResNet study also introduced the use of bottleneck blocks which stacks three convolution blocks of 1*1, 3*3, and 1*1 kernel respectively instead of 2 layers for deeper networks. Parameter-less identity shortcuts showcased better performance in bottleneck-based ResNet architectures. The 152 Layer ResNet replaced 2-layer blocks of Resnet-34 using 3-layer bottlenecks which resulted in better accuracy in the ImageNet challenge. ResNet152V2 is described by \cite{b13} as a pre-activated rectified linear unit (RELU) with the original ResNet152.

\subsection{MobileNetV2}

MobileNet focused on building lightweight and efficient visual domain deep neural networks introduced by \cite{b14}. Its core was based on depth-wise separable convolutions, which resulted in ~9 times less computation than standard convolutions while sacrificing only a small amount of accuracy. The width and resolution multiplier hyperparameters are intended to optimize the tradeoff between classifier latency and accuracy. MobileNetV2 was introduced next year by \cite{b23} and designed with an inverted residual structure with the shortcut connections between the thin bottleneck layers. MobileNetV2 was found to be a significant upgrade from MobileNetV1 and demonstrates state of art results for some near-real-time visual classification domain problems. MobileNetV2 brought in skip connections between the newly introduced linear bottleneck between layers in the architecture. The experiments found that the MobileNetV2 required half the operations and 30\% lesser parameters than MobileNetV1 while achieving better accuracy. 

\subsection{Transfer Learning}

Transfer learning was conceptualized by taking inspiration from the human learning process. Human society earns knowledge by transferring it generation by generation and then adding on to it. Transfer learning is a concept where one model built for a specific task is reused on a different task. One of the major challenges of artificial intelligence implementation is the availability of quality tagged datasets. In the deep learning world, training robust and scalable models takes a huge amount of computing resources for image and text data considering the relevant dataset availability. Commonly used transfer learning tasks are to select a pre-trained model suitable for the task, reuse the model entirely without any weight updates, and choose to tune the model by updating the weight for the last few layers. Transfer learning can be thought of as an optimization mechanism that saves time, cost, and resources. Transfer learning models also can result in better accuracy than a model built from scratch as transfer learning models have learned feature representations from the available tagged categorical dataset and are usually able to generalize better. This study will experiment with transductive learning where pre-trained models are from the visual domain, but the target application domain is cardiac sound.

\subsection{Proposed methods}

Several pre-defined feature extractors experimented with this study with different configurations for classification requirements. Approaches such as using the existing model without any weight enhancement, updating weights in the last few layers, and tuning weights of all layers were evaluated during the experiments.

\begin{figure*}[htbp]
{\includegraphics[width=500pt]{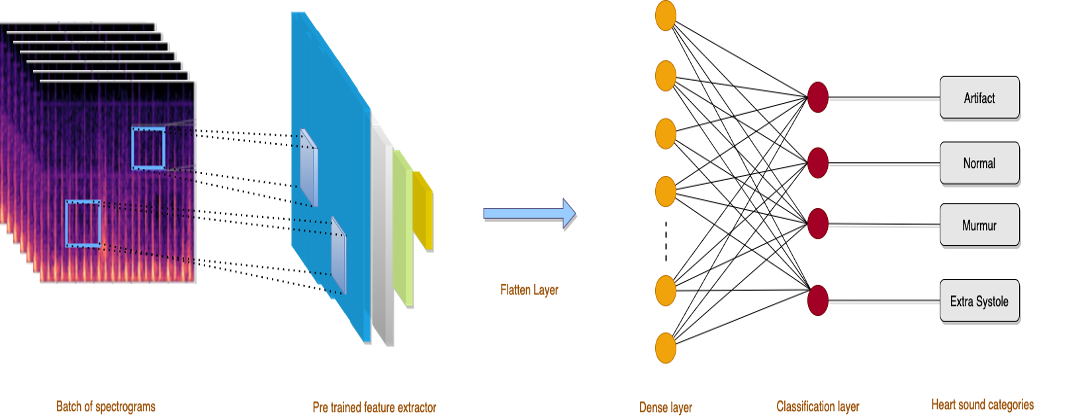}}
\caption{Cardiovascular audio classification transfer learning layers}
\label{fig}
\end{figure*}

Some common configurations for all models were used across the training process. All the experiments operated with a consistent input size of 128*128 over RGB channels. Different values such as 224*224, 164*164, and 100*100 also experimented but none of them performed better than the chosen dimension. The batch size was kept as 8 to ensure the training process did not go out of memory. Keras image data generator method was applied to supply data on runtime while training and ensure to shuffle the data for each epoch. The last layer was ignored for all pre-defined feature extractors used for transfer learning in this study to ensure the learned feature maps are only used during cardiovascular audio classification. On all optimizer configurations learning decay rate has been enabled dynamically based on the defined learning rate and epochs. Decay value was defined as the learning rate divided by the number of epochs for this study in all the experiments. All the predefined feature extractor architectures were loaded with ImageNet initialized weights. Previously calculated class weights had been assigned to all categories to ensure weighted loss calculation during the training process. Tensorflow based ‘Keras’ python library was used for training, validation, and testing purposes.

The pre-trained networks extract meaningful and detailed feature maps which are then required to be used for further classification. These feature maps are required to be first flattened without impacting the batch size. For example, in case the featured maps are with a dimension of (64 * 10 * 10) where 64 is the batch size and the shape of the input is 10*10, then the flatten layer will output will be of size 64*100. The input and output size of the flatten layer will vary depending upon the pre-trained network used. To stop a neural network to overfit several regularization techniques can be used with a neural network such as L1, L2, DropOut, etc. This study used dropout and batch normalization layers between both the fully connected layers wherever the training resulted in an overfitted model. Not all the experiments included these layers and were used as a need basis. A dropout value of 0.2, 0.3, 0.35, etc experimented with different architectures. Dense layers have been used to process input from the flatten layer and then transform the them with activation and calculated weights and biases. These transformed signals were then passed to another dense fully connected layer for categorization between cardiovascular audio for both the dataset. The first fully connected layer used rectified linear activation function or ReLU as activation function and the second layer used SoftMax as activation. First dense layer configured to experiment with 32, 64, 128, 500, and 1000 neurons. Datasets A and B have 4 and 3 categories of cardiovascular audio data respectively. Hence the number of neurons in the final layer was selected as 4 and 3 respectively for the dataset A and B based on the output categories. 

Optimizers experimented with the study will be stochastic gradient descent (SGD) and ADAM. Momentum value in case of SDG optimizer is considered as 0.9 throughout. Learning rates experimented are 0.01, 0.001, and 0.0001. Number of epochs assessed in a range between 10 to 40. The study has used categorical cross-entropy as the loss function as the cardiovascular audio category detection is a multi-class classification problem. Categorical cross-entropy loss is a strong indicator of how easily two discrete probability distributions can be distinguished from one another. As the distributions get closer to each other the loss gets smaller. Mathematically the categorical cross-entropy is resolved to be with equation 7 where y is the actual label and p is the predicted label. The loss function requires the output to be defined as one-hot vectors, hence the class labels are required to be converted accordingly. The activation function on the final layer is a SoftMax function to support categorical cross-entropy as a loss function. 

\begin{equation}
-y^{T} . \log (p)=-\left[\begin{array}{lll}
y_{1} & y_{2} & y_{3}
\end{array}\right] \cdot\left[\begin{array}{l}
\log \left(p_{1}\right) \\
\log \left(p_{2}\right) \\
\log \left(p_{3}\right)
\end{array}\right]
\end{equation}

\begin{figure*}[]
{\includegraphics[width=500pt,height=250pt]{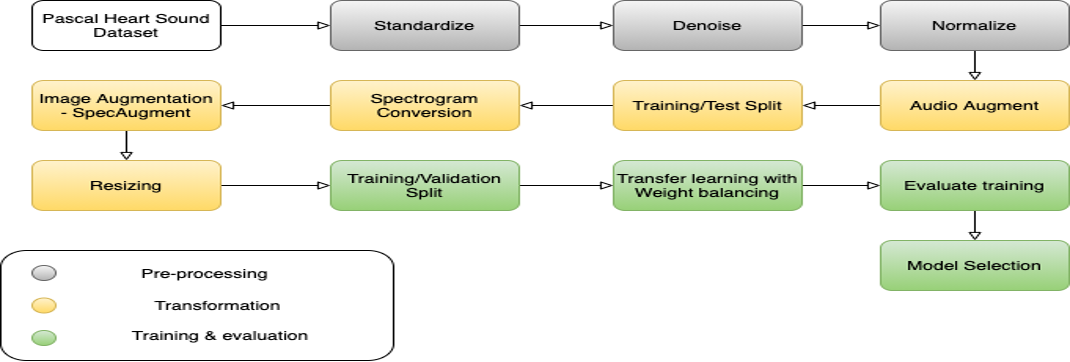}}
\caption{Detailed level process of Cardiac audio classification}
\label{fig}
\end{figure*}

A callback can help to automate certain tasks like saving the best model based on selected criteria, stopping training early based upon the value of certain metrics, or reducing the learning rate in case of no training improvements. The current study persists the model when the training process notices validation loss is minimum for already executed iterations. The ReduceLROnPlateau function of python Keras library helps to wait till the execution of a defined number of epochs with no improvement with chosen metrics and then reduce the learning rate by a declared factor. In this study, the patience level was defined as 5 epochs and the learning rate reduction factor was 0.2. The metrics to monitor the model performance were defined as validation loss.

\section{RESULTS AND ANALYSIS}

Experiments were conducted separately for datasets A and B which also will be compared separately based on key performance indicators. Different pre-trained visual domain convolutional neural network architectures with some updates in the last few layers will be evaluated based on the training, validation, and testing performance. Generated spectrograms for different heart sound categories found to be visually separable. The upper half of the spectrograms are mostly black as the higher range of the frequency ranges were filtered out with the denoising process. The results will also be analyzed to understand the best-suited model for cardiovascular audio classification. 

\begin{figure}[]
{\includegraphics[width=350pt,height=200pt]{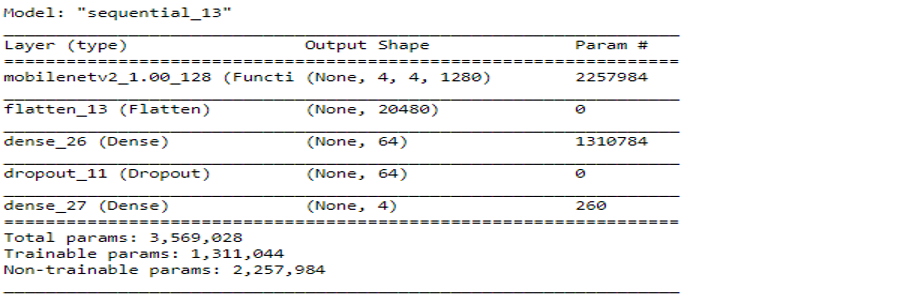}}
\caption{summary of a sample model used for cardiovascular audio classification}
\label{fig}
\end{figure}

\begin{figure*}[h!]
{\includegraphics[width=500pt,height=250pt]{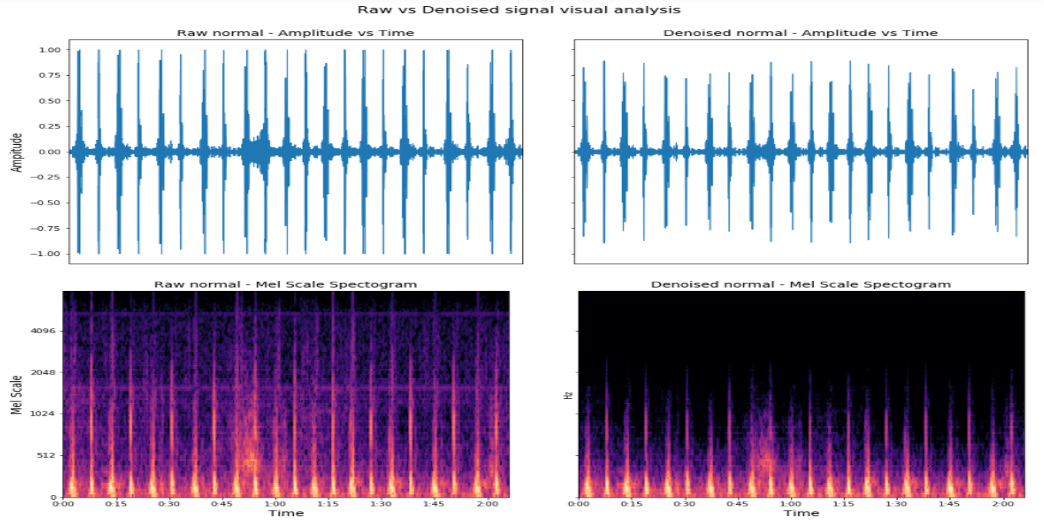}}
\caption{Raw vs Denoised signal analysis}
\label{fig}
\end{figure*}

\begin{figure*}[]
{\includegraphics[width=500pt]{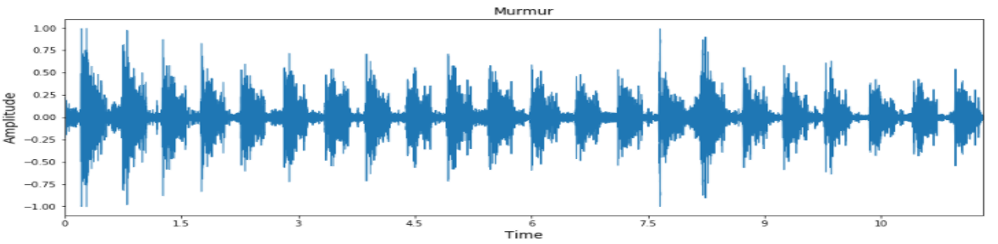}}
{\includegraphics[width=500pt]{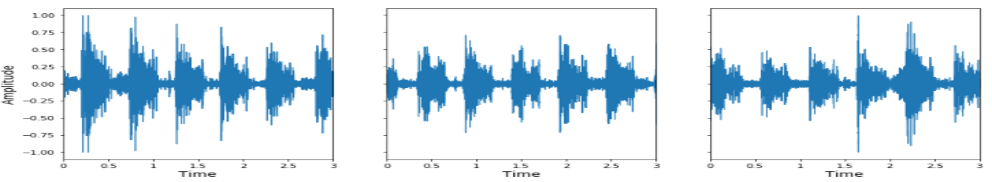}}
\caption{A 11 seconds long cardiac sound split each 3 seconds chunk, last 2 seconds ignored}
\label{fig}
\end{figure*}

\begin{figure*}[h!]
{\includegraphics[width=500pt,height=250pt]{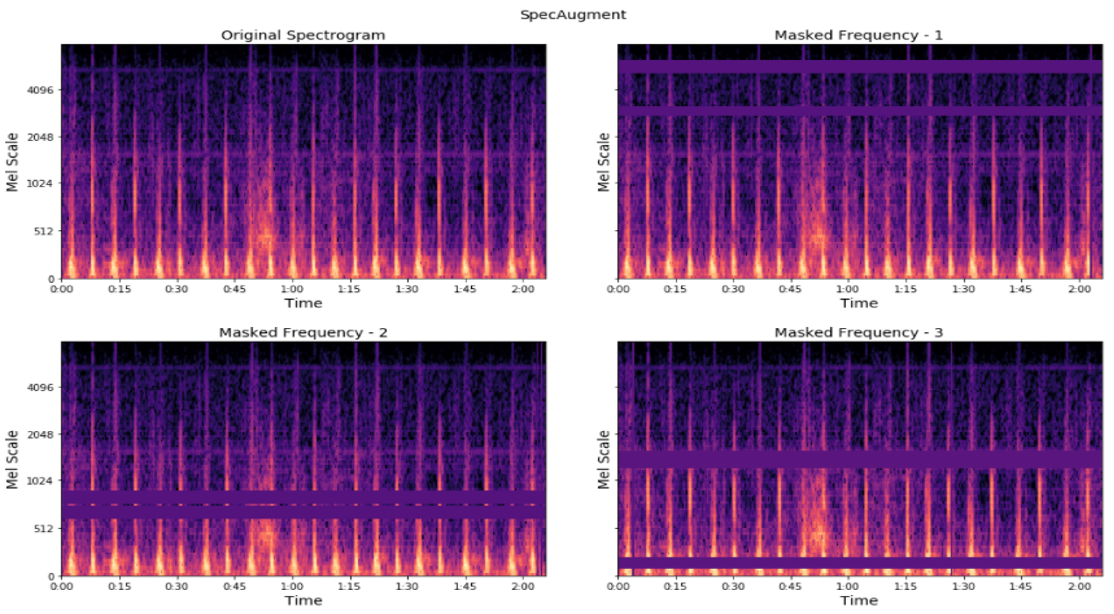}}
\caption{Spectrogram Augmentation – 3 frequency masked spectrograms along with the original}
\label{fig}
\end{figure*}

Audio files split mechanism was successfully implemented, for example, if a record is of 11 seconds in length, then it was split into 3 parts i.e. 1 – 3 seconds, 4 - 6 seconds, and 7 - 9 seconds. The part that includes the 10th and 11th seconds was ignored; example presented in figure 5. Spectrogram augmentation technique has helped to boost the number of available images for better training of applicable visual domain neural networks. This study experimented with batch normalization with different layers but did not find any improvement with the model performance, hence omitted in the final experiments. Most of the models resulted in a good performance with 64 neurons for the second last dense layer along with the rest of the architecture, hence the study chooses the same number of neurons for final experiments.

\subsection{Results of experiments on Dataset A}

The sampling frequency of the captured audio for dataset A is 44100 Hz. High sample frequency like this can capture accurate audio information and result in larger file size. During neural network training, a batch of these data will be loaded directly into memory for classification requirements. This caused out-of-memory issues during the training process. All the cardiovascular audio samples of dataset A were downsampled by a factor of 10 to reduce the file size and information distribution. Operations with newly generated files with sampling frequency 4410 Hz were faster and no memory inconsistency issue was observed. The dataset is small and only consists of 291 samples after the split whereas 41\% of samples belong to the artifact category. It was preprocessed with all the steps mentioned earlier except the audio augmentation step. Some of the sample models were then trained with the generated and augmented spectrogram images. Pre-trained neural networks i.e. MobileNet, MobileNetV2, ResNet152V2, etc. were tuned based on updated class weights. The training process was found to be fast due to the availability of less data, but classification performance was not satisfactory. The highest test categorical accuracy value resulted in 0.81, precision as 0.84, and recall value as 0.74 except audio augmentation.

\begin{figure}[]
{\includegraphics[width=250pt,height=175pt]{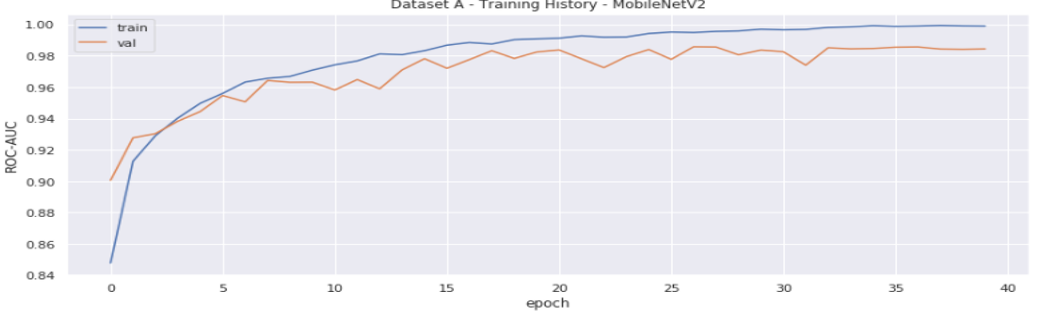}}
\caption{MobileNetV2 training history W.R.T AUROC for Dataset A}
\label{fig}
\end{figure}

Audio augmentation techniques were then implemented with the selected training dataset which generated 2x added audio files with pitch and time shift. The training dataset contained ~1500 samples and the validation set contained ~500 samples after the audio augmentation. ResNet152V2 is the largest pre-trained network tested with cardiovascular audio in this study. Added with post flattening and classification layers the network was trained with a stochastic gradient descent optimizer. The best key performance indicators with this network were achieved when the learning rate and momentum values were set to 0.001 and 0.9 respectively. A dropout value of 0.35 was employed as regularization step while the inputs  were transmitted from the previous dense layer with 64 neurons to the last classification layer. The training process continued till 40 epochs where the 35th epoch resulted in the best outcome. Test data evaluation showed the test loss as 0.3942, categorical accuracy as 0.9041, precision as 0.9041, recall as 0.9041, and AUROC as 0.9797.

MobileNet network as pre-trained layer followed by flattening and two dense layers also resulted in good classification outcome. It implemented Adam optimizer with 0.001 learning rate and reusing existing weights for the ImageNet pre-trained MobileNet network. No dropout or batch normalization has been used during the model training. The experiment was configured for 15 epochs, but the best result was captured during the 12th epoch. Resulted test metrics were test loss as 0. 5142, categorical accuracy as 0.8767, precision as 0.8767, recall as 0.8767, and AUROC as 0.9629. ImageNet pre-trained MobileNetV2 network along with two fully connected layers experimented with similar settings as the MobileNet network. The finest model was found where the training process was performed with a stochastic gradient descent optimizer, 0.001 leaning rate, and 0.9 as momentum value. As a regularization step, a drop-out value of 0.3 was used while the inputs were passed to the last classification layer from the previous dense layer with 64 neurons. The best result was observed when the existing pre-trained weights were not updated during the training process for dataset A. 40 epochs were set up with the mentioned configuration and the top outcome was found at the 33rd epoch. Test loss was captured as 0.4516, categorical accuracy as 0.8767, precision as 0.8873, recall as 0.863, and AUROC as 0.9585. Training time required was less in MobileNet than MobileNetV2 due to the fewer epochs. The training history of the MobileNetV2 based model concerning the AUROC score can be found in figure 8. 

\begin{figure}[]
{\includegraphics[width=250pt,height=175pt]{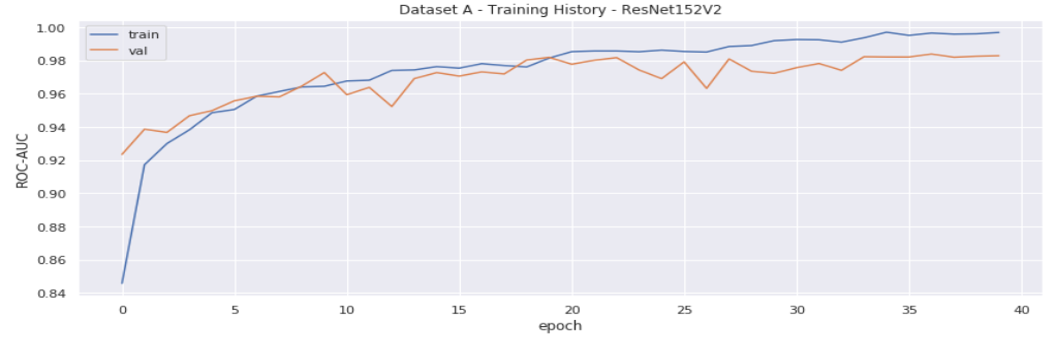}}
\caption{ResNet152V2 training history W.R.T AUROC for Dataset A}
\label{fig}
\end{figure}

Several other pretrained networks, i.e. InceptionResNetV2, Xception, and DenseNet169 with varied settings were evaluated, the result is available in table II. Experiments show the ResNet152V2 based model to be the best classifier for dataset A with the highest AUROC, categorical accuracy, precision, and recall. MobileNet and MobileNetV2 based models also showed reliable classification performance metrics for the dataset. All three models did not fine-tune any of the pre-trained weights and reused the ImageNet weights for feature extraction of spectrogram images. Hence, it can be concluded that the image domain CNN based models can extract suitable low and high-level features from the audio domain spectrograms which can then be used for cardiovascular sound categorization successfully. The training history of the ResNet152V2 based model concerning the AUROC score can be found in figure 9.

\begin{table*}[t!]
\caption{ \label{tab:table-name}  Experiment results with Dataset A}

\begin{tabular}{|l|l|l|l|l|l|}
\hline
\multicolumn{2}{|l|}{\textbf{Dataset - A}}                                                                                                                                                                                                                                       & \multicolumn{4}{l|}{\textbf{Test results}}                                                                                     \\ \hline
\textbf{Feature Extractor}                                                            & \textbf{Model summary}                                                                                                                                                                   & \textbf{\begin{tabular}[c]{@{}l@{}}Categorical \\ Accuracy\end{tabular}} & \textbf{Precision} & \textbf{Recall} & \textbf{AUC} \\ \hline
\begin{tabular}[c]{@{}l@{}}MobileNet –\\    \\ Pretrained weights\end{tabular}        & \begin{tabular}[c]{@{}l@{}}1 Flatten, 1 dense with 64 neurons, \& 1 classification   layer. \\ Adam Optimizer with LR 0.001, Best result – 12th Epoch\end{tabular}                       & 0.8767                                                                   & 0.8767             & 0.8767          & 0.9629       \\ \hline
\begin{tabular}[c]{@{}l@{}}MobileNetV2 –\\    \\ Pretrained weights\end{tabular}      & \begin{tabular}[c]{@{}l@{}}1 Flatten, 1 dense with 64 neurons, dropout(0.3)  \& 1 classification layer. \\ SGD Optimizer   with LR 0.001, Best result – 33rd Epoch\end{tabular}          & 0.8767                                                                   & 0.8873             & 0.863           & 0.9585       \\ \hline
\begin{tabular}[c]{@{}l@{}}InceptionResNetV2-\\    \\ Fine-tuned weights\end{tabular} & \begin{tabular}[c]{@{}l@{}}1 Flatten, 1 dense with 64 neurons \& 1 classification layer.   \\ SGD Optimizer with LR 0.01, \& M 0.9, Best result – 6th Epoch\end{tabular}                 & 0.8493                                                                   & 0.8451             & 0.8219          & 0.9217       \\ \hline
\begin{tabular}[c]{@{}l@{}}ResNet152V2 –\\    \\ Pretrained weights\end{tabular}      & \begin{tabular}[c]{@{}l@{}}1 Flatten, 1 dense with 64 neurons, dropout(0.35)  \& 1 classification layer. \\ SGD Optimizer   with LR 0.001\& M 0.9, Best result – 35th Epoch\end{tabular} & 0.9041                                                                   & 0.9041             & 0.9041          & 0.9797       \\ \hline
\begin{tabular}[c]{@{}l@{}}Xception –\\    \\ Fine-tuned weights\end{tabular}         & \begin{tabular}[c]{@{}l@{}}1 Flatten, 1 dense with 64 neurons, \& 1 classification   layer. \\ Adam Optimizer with LR 0.001, Best result – 6th Epoch\end{tabular}                        & 0.7945                                                                   & 0.8182             & 0.7397          & 0.9573       \\ \hline
\begin{tabular}[c]{@{}l@{}}DenseNet169 –\\    \\ Pretrained weights\end{tabular}      & \begin{tabular}[c]{@{}l@{}}1 Flatten, 1 dense with 64 neurons, \& 1 classification   layer. \\ Adam Optimizer with LR 0.001, Best result – 12th Epoch\end{tabular}                       & 0.863                                                                    & 0.863              & 0.863           & 0.9499       \\ \hline
\end{tabular}
\end{table*}

\subsection{Results of experiments on Dataset B}

The cardiovascular sounds for dataset B were recorded with a sampling frequency of 4000 Hz. This further did not require any resampling as the sampling frequency is already on the lower side and provides enough information for classification. Dataset B consists of a total of 758 samples after the initial split of each audio file into 3 seconds chunks. ResNet152V2 along with the flattening and classification layers was trained by fine-tuning the weights of the pre-trained network to attain the finest classification metrics for dataset B. The model was trained with Adam optimizer, 0.001 as the learning rate, and achieved the best results during the 15th epoch. Adding dropout or batch normalization did not showcase any improvement in the training performance. Test results were evaluated with unseen test data and resulted in test loss as 0.3697, categorical accuracy as 0.899, precision as 0.9069, recall as 0.8894, and AUROC as 0.9636. MobileNetV2 embedded network training process was performed using stochastic gradient descent optimizer along with 0.001 learning rate and 0.9 value to achieve the top results for defined metrics. This was configured to be trained for 10 epochs and the best results were found on the 7th epoch. The finest outcome was achieved by the model which used fine-tuning weights of pre-trained MobileNetV2 network. The best results achieved for the MobileNetV2 based network are test loss as 0.5691, categorical accuracy as 0.8942, precision as 0.8942, recall as 0.8942, and AUROC as 0.9537.

\begin{figure}[]
{\includegraphics[width=250pt,height=175pt]{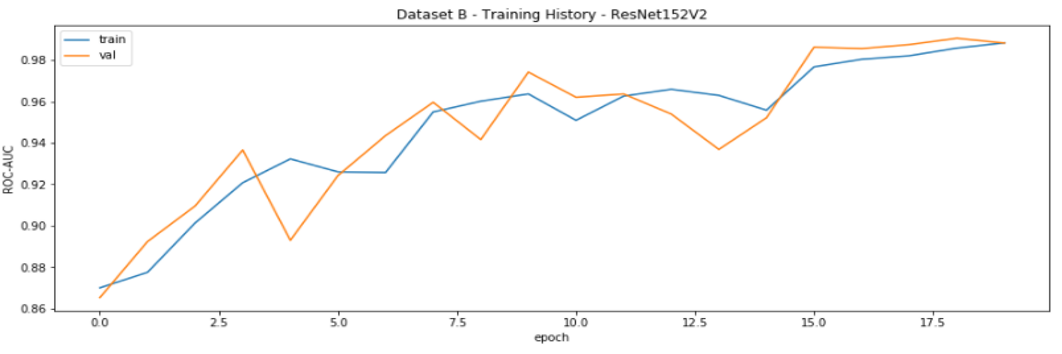}}
\caption{ResNet152V2  training history W.R.T AUROC for Dataset B}
\label{fig}
\end{figure}

Other evaluated pre-trained feature extractors were MobileNet, InceptionResNetV2, and Xception along with the classification network. All conducted experiments conclude that the ResNet152V2 based model performs best for the classification of dataset B. Test data evaluation with this scored an AUROC of 0.96 and categorical accuracy of 0.9069. Both MobileNetV2 and Xception based networks can be considered as good classifiers of dataset B as they scored ~0.95 with AUROC and 0.894 with categorical accuracy. All the experiments with dataset B resulted better when all the layers of pre-trained networks were fine-tuned appropriately. This may be because of weight-based loss calculation while the training process as the dataset includes significant class imbalance. Table 3 shows the validation and test scores of all the experimented pre-trained visual domain-based neural networks with dataset B.

\begin{table*}[t]
\caption{ \label{tab:table-name}  Experiment results with Dataset B}

\begin{tabular}{|l|l|l|l|l|l|}
\hline
\multicolumn{2}{|l|}{\textbf{Dataset - B}}                                                                                                                                                                                                                           & \multicolumn{4}{l|}{\textbf{Test results}}                                                                                     \\ \hline
\textbf{Feature Extractor}                                                            & \textbf{Model summary}                                                                                                                                                       & \textbf{\begin{tabular}[c]{@{}l@{}}Categorical \\ Accuracy\end{tabular}} & \textbf{Precision} & \textbf{Recall} & \textbf{AUC} \\ \hline
\begin{tabular}[c]{@{}l@{}}MobileNet –\\    \\ Fine-tuned weights\end{tabular}        & \begin{tabular}[c]{@{}l@{}}1 Flatten, 1 dense with 64 neurons, \& 1 classification layer.  No Dropouts.\\ Adam Optimizer with LR 0.01, Best result – 15th Epoch\end{tabular} & 0.899                                                                    & 0.899              & 0.899           & 0.9382       \\ \hline
\begin{tabular}[c]{@{}l@{}}MobileNetV2 –\\    \\ Fine-tuned weights\end{tabular}      & \begin{tabular}[c]{@{}l@{}}1 Flatten, 1 dense with 64 neurons, \& 1 classification layer. No Dropouts.\\ SGD Optimizer with LR 0.001 \& M 0.9, Best result – 7th   Epoch\end{tabular}    & 0.8942                                                                   & 0.8942             & 0.8942          & 0.9537       \\ \hline
\begin{tabular}[c]{@{}l@{}}InceptionResNetV2-\\    \\ Fine-tuned weights\end{tabular} & \begin{tabular}[c]{@{}l@{}}1 Flatten, 1 dense with 64 neurons, \& 1 classification layer. No Dropouts.\\ SGD Optimizer with LR 0.001 \& M 0.9, Best result – 5th   Epoch\end{tabular}    & 0.8942                                                                   & 0.8942             & 0.8942          & 0.951        \\ \hline
\begin{tabular}[c]{@{}l@{}}ResNet152V2 –\\    \\ Fine-tuned weights\end{tabular}      & \begin{tabular}[c]{@{}l@{}}1 Flatten, 1 dense with 64 neurons, \& 1 classification layer. No Dropouts.\\  Adam Optimizer with LR 0.001, Best result – 15th Epoch\end{tabular}             & 0.899                                                                    & 0.9069             & 0.8894          & 0.9636       \\ \hline
\begin{tabular}[c]{@{}l@{}}Xception –\\    \\ Fine-tuned weights\end{tabular}         & \begin{tabular}[c]{@{}l@{}}1 Flatten, 1 dense with 64 neurons, \& 1 classification layer. No Dropouts.\\ Adam Optimizer with LR 0.001, Best result – 5th Epoch\end{tabular}              & 0.8942                                                                   & 0.8932             & 0.8846          & 0.9587       \\ \hline
\end{tabular}
\end{table*}

\section{MODEL EVALUATION AND DISCUSSION}

The experiment results demonstrate the applicability of visual domain neural networks with the cardiovascular audio dataset, primarily because the pre trained image domain knowledge was able to identify low and high level features of heart sound spectrograms well. Splitting each audio file in multiple 3-second chunks helped to retain useful cardiovascular sound information and increased the number of samples available for classification. Usage of spectrogram augmentation technique was also demonstrated to be successful with cardiovascular audio datasets that boosted the classification accuracy and AUROC scores. Audio augmentation techniques like time and pitch shift to boost the number of audio samples showed to be very effective for the classification results of dataset A.

The ResNet152V2 based feature extractors were found to be the top identifier of characteristics for both datasets with an classification network AUROC score of  0.9797 and 0.9636 respectively for datasets A and B. ImageNet pre-trained weights for all layers were used without any change for classification of dataset A whereas weights of all layers were fine-tuned during the training process for dataset B. The family of MobileNet based feature extractors also showcased good ability to recognize detailed features with both datasets. As MobileNet based feature extractors are small and very fast during the training process, they can be used in mobile devices for near-real-time interpretation and feature extraction. MobileNetV2 was found to be more stable in characteristics identification the classes with a classification network AUROC score of 0.9585 and 0.9537 for datasets A and B respectively. Similar to ResNet152V2, the pre-trained ImageNet weights are used for all layers, and there is no change to classify dataset A, while the weights of all layers are adjusted when training dataset B. This phenomenon can be related to the degree of class imbalance between datasets. The training process for dataset B included class weight balancing and weight-wise cross-entropy loss calculation. This can be the reason why fine-tuning of all layers resulted in better model performance for all networks experimented with Dataset B.

Most of the other experimented networks resulted in good classification metrics as well. i.e. DenseNet for dataset A and XceptionNet for dataset B. Results prove the effectiveness of the approach of transferring the visual domain knowledge into audio domain data classification. ImageNet learned weights helped to extract required cardiovascular audio-specific features like identifying peaks, murmur noise, extra sound peak, etc which in turn resulted in better classification metrics score. The approach experimented with the study can be used universally to train and classify heart sounds with good categorical accuracy, precision, AUROC, etc.  

\footnote{https://github.com/uddipanmukherjee/cardiovascular\_audio\_classification}

\section{CONCLUSIONS AND RECOMMENDATIONS}

ResNet152V2 based feature extractor has 152 layers and ~60.3 million parameters making it one of the large ImageNet trained pre-trained networks. Even though the categorical accuracy and AUROC score are higher, the inference time was also noted to be higher. MobileNetV2 only consists of ~3.5 million parameters making it one of the fastest feature extractors in the experimented pre-trained networks. Classification networks used features extracted with MobileNetV2 also obtained a good AUROC for both datasets A and B, hence it can be utilized for classification for near real-time inference requirements. However, ResNet152V2 based feature extractor with the following classification network is recommended to be used for inference whenever classification accuracy is more important than the time required for inference.

\subsection{Knowledge contribution}

In previous studies domain-based feature extraction techniques have been used to manually identify appropriate features for cardiovascular datasets. To our knowledge, none of the research has explored image domain transferred automated feature extraction and classification for cardiovascular audio datasets. This study successfully demonstrates how visual domain neural networks can be used as feature extractors to categorize heart audio data with good accuracy. The method described in this study can be further used to train on other cardiovascular datasets and predict high-level categories. ResNet152V2 and MobileNetV2 based feature extractors along with following classification layers were found to be good classifiers of cardiovascular datasets presented in the Pascal heart sound classification challenge.

State of the art spectrogram augmentation method SpecAugment experimented with speech data and showed good speech recognition results. As per our knowledge, this study has used the specAugment method to cardiovascular audio data for the first time. Implementation of the method augmented the heart sound spectrograms effectively by time wrapping and frequency masking techniques and overall boosted the classification accuracy.

\subsection{Study Limitations}

The approach used in this study can be used universally to train and classify any heart sound dataset targeting to categorize labels such as normal, murmur, extrasystole, extrahls, etc. The consideration for this study was that the cardiovascular audio was visually distinguishable for these categories with help of mel scale spectrograms. Now the proposed method in this study may not hold in case any categorization of heart sounds is not visually distinguishable. The study also did not identify the exact location of S1, S2 peaks, or the murmur sound to classify heart sounds. The approach discussed in the study may not help to identify such locations in the given cardiovascular audio.

\subsection{Future Work}

The study proposed a simplistic approach for denoising the audio files by using a low pass filter to eliminate any sound above 192 Hz. This might have removed some murmur and heart sounds as well, also there can be scenarios where external noise of low frequency was mixed with the heart sound below the threshold frequency. A denoising method can be worked upon to take care of these complex scenarios, which can result in suitable denoised audio for classification. Dataset A and B consist of varying audio lengths between 1 and 30 seconds. This study chose 3 seconds as a standard for splitting each audio file as a full cycle of heart sound should be covered within that time frame. Any audio length less than 3 seconds was discarded as it does not contain enough information required for classification. As future work, this duration can be increased and experimented with other values if that impacts the classification performance.

The current study implements audio augmentation techniques with dataset A, and spectrogram augmentation techniques with both datasets. As a future recommendation, audio augmentation techniques like time and pitch shift can be tested with dataset B. SpecAugmentation could be used to generate more frequency masked spectrograms per original spectrogram to check if it helps to boost the classification performance. Audio augmentation techniques like time stretch and volume gain could be applied to both datasets. Another future recommendation is to increase the input image size from 128*128 to 224*224 to enable better feature extraction. Preprocessing steps and the training process need to be performed in a high memory configured GPU environment to achieve this.

\bibliographystyle{ieeetr}

\end{document}